\documentclass[english,showpacs,floatfix]{revtex4}
\usepackage[T1]{fontenc}
\usepackage[latin1]{inputenc}
\usepackage{amsmath}
\usepackage{amssymb}
\usepackage{graphicx}

\makeatletter
\@ifundefined{textcolor}{}
{%
 \definecolor{BLACK}{gray}{0}
 \definecolor{WHITE}{gray}{1}
 \definecolor{RED}{rgb}{1,0,0}
 \definecolor{GREEN}{rgb}{0,1,0}
 \definecolor{BLUE}{rgb}{0,0,1}
 \definecolor{CYAN}{cmyk}{1,0,0,0}
 \definecolor{MAGENTA}{cmyk}{0,1,0,0}
 \definecolor{YELLOW}{cmyk}{0,0,1,0}
}

\usepackage{longtable}\usepackage{dcolumn}\usepackage{babel}

\makeatother

\usepackage{babel}
\begin{document}

\title{Regular rotating black holes and the weak energy condition}

\author{J. C. S. Neves}

\email{nevesjcs@ime.unicamp.br}

\author{Alberto Saa}

\email{asaa@ime.unicamp.br}

\affiliation{Departamento de Matem\'atica Aplicada, Universidade Estadual de Campinas \\
13083-859 Campinas, SP, Brazil}
\begin{abstract}
We revisit here a recent work on regular rotating black holes. 
We introduce a new mass function generalizing 
the commonly used Bardeen and Hayward mass functions and  extend 
the recently proposed solutions in order to accommodate  a 
cosmological constant $\Lambda$.
We discuss some aspects of the causal structure (horizons) and the ergospheres of the 
new proposed solutions. 
 We also show that, in contrast with the spherically symmetrical case, 
the black hole rotation will unavoidably lead to the violation of the weak energy condition for any physically reasonable
choice of the mass function, reinforcing the idea the description of the interior region of a Kerr black hole
is much more challenging than in   the Schwarzschild case. 
\end{abstract}


\maketitle

\section{Introduction}

The problem of spacetime singularities is an open issue in Physics (see, for instance, \cite{sing} for a general
discussion and \cite{Novello} for some cosmological implications). The general and commonly accepted  belief is that only a not yet
available quantum theory of gravity would be capable of solving them properly. In recent years, without a fully developed  and reliable 
candidate for
a quantum theory of gravity, many phenomenological models have been proposed for which the central 
singularity of a black hole is avoided (see, for a review and motivations, \cite{Ansoldi}). These non-singular solutions of General Relativity
 are the so-called regular black holes (BH) and,  since there are strict 
uniqueness theorems for BH solutions of vacuum Einstein-Maxwell equations \cite{Heusler}, they will necessarily  require some kind of exotic matter/field or internal structure in order to exist. The typical stationary spherically symmetrical regular BH has line element
\begin{equation}
ds^2 = - f(r)dt^2 + \frac{dr^2}{f(r)} + r^2d\Omega^2,
\end{equation}
where  $d\Omega^2=d\theta^2+\sin^2\theta d\phi^2$ and $f(r) = 1 - {2m(r)}/{r}$. A mass function of the type 
\begin{equation}
\label{massfunc}
m(r) = \frac{M_0}{\left(1 + \left(\frac{r_0}{r}\right)^q \right)^\frac{p}{q}}
\end{equation}
will 
guarantees an asymptotic flat spacetime for positive $p$ and $q$. $M_0$ and $r_0$ are, respectively, a mass and a length parameters. 
The well known Bardeen \cite{Ansoldi} and Hayward \cite{Hayward}  BH correspond, respectively, to the choices $p=3$, $q=2$ and $p=q=3$ 
in the mass function expression (\ref{massfunc}). The limits of small and large $r$ of
(\ref{massfunc}) are, respectively,
\begin{equation}
\label{small}
m(r) \approx M_0 \left(\frac{r}{r_0} \right)^p
\end{equation}
and
\begin{equation}
\label{large}
m(r) \approx M_0 \left( 1 - \frac{p}{q}\left(\frac{r_0}{r} \right)^q\right).
\end{equation}
Both Bardeen and Hayward BH are realizations of a quite old idea, introduced by Sakharov and co-workers in the sixties \cite{Sakharov} and
later improved \cite{Markov},
that spacetime in the highly dense central region of a BH would be de-Sitter like
\begin{equation}
\label{desitter}
G_{\mu\nu} = -\lambda g_{\mu\nu},
\end{equation}
with $\lambda >0$, which requires 
\begin{equation}
\label{r=0}
f(r) = 1 - \left(\frac{r}{\ell}\right)^2
\end{equation}
for $r\approx 0$. 
Comparing with (\ref{small}), we see that (\ref{r=0}) demands $p=3$. 
The only typical requirement on $q$ is to get from (\ref{large}) an asymptotically Schwarzschild solution, what requires only $q>0$.
We will consider here the general mass function with $p=3$ and $q>0$. Figure 1 depicts some typical cases. 
\begin{figure}
\begin{centering}
\includegraphics[scale=0.57]{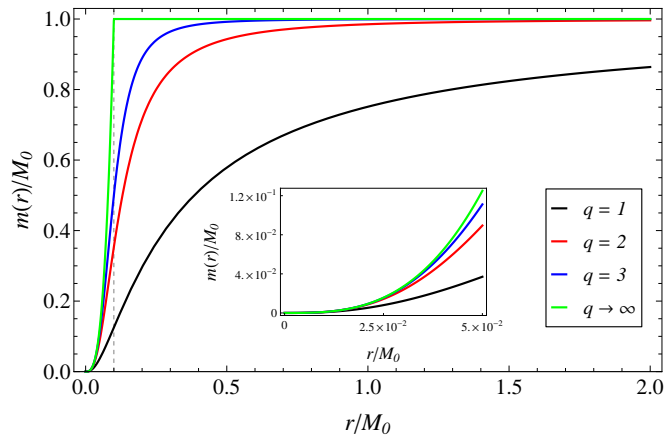}
\par\end{centering}
\caption{Mass function (\ref{massfunc}) with $p=3$ and $q>0$. The parameter $r_0$ is typically assumed to
be microscopic ($r_0 \ll M_0$) and, hence, the exterior region of the BH can be very close to the Schwarzschild
spacetime. The $q\to\infty$ case corresponds to the usual matching between de Sitter and Schwarzchild solutions in the
interior region of the BH. The dashed line indicates $r=r_0$. In this graphic we have used  $r_0/M_0=10^{-1}$.}
\end{figure}
Notice that for the mass function $(\ref{massfunc})$ with $p=3$
one has
\begin{equation}
\lambda = {\frac{6M_0}{r_0^3}}. 
\end{equation}
Since $r_0$ is typically assumed to be microscopic ($r_0 \ll M_0$), the core density described by the central de Sitter solution (\ref{desitter}) 
may be effectively very high, possibly in the regime where quantum gravity effects should come out. For the spherically symmetrical case (no rotation), several mass functions interpolating between
the de Sitter core ($r\approx 0$) and the asymptotically flat infinity  ($r\to\infty$) lead to physically reasonable regular 
black holes since, despite of violating the strong energy condition as required by the singularities theorem \cite{sing}, they do obey
the weak energy condition  and, hence, might be in principle formed from a physically reasonable matter content.

In the recent work \cite{Bambi}, Bambi and Modesto explore the  Newman-Janis algorithm \cite{Newman-Janis}
  in order to construct rotating regular BH with Bardeen and Hayward
mass functions. One of their conclusions is that for these two commonly used mass functions, the weak energy condition (WEC)
is violated due to the rotation of the black hole. Despite of being physically problematic due to the violations of WEC and, hence, to the
 presence of  negative energy density content somewhere, such solutions as those ones introduced by Bambi and Modesto are certainly
interesting from a phenomenological point of view, since astrophysical bodies, the main data sources for exploring BH physics, typically
have nonvanishing angular momentum.
 In the present paper, we extend the Bambi and Modesto solutions for the case
where a cosmological constant $\Lambda$ is present, a situation which could be useful, for instance, to the studies involving rotating black holes and the AdS/CFT conjecture \cite{Hawking}. We consider 
 mass functions of the type (\ref{massfunc}) with $p=3$ and $q>0$, but some of our conclusions are valid for any
 physically reasonable functions, {\em i.e.}, functions compatible with the behavior (\ref{desitter})-(\ref{r=0}) near the origin.
 We discuss some aspects of the causal structure (event, cosmological, and Cauchy horizons) and the ergospheres of the new proposed solutions. We show also that
the violation of WEC is indeed generic and unavoidable for rotating BH, irrespective of the used mass terms, with the only requirement that
they behave as $m(r)\propto r^3$ for $r\to 0$, which is  necessary to have a behavior similar to
(\ref{desitter}) and hence to have an extremely dense central region, but free of 
  singularities. Our result is another indication  that the description of the interior region of the Kerr BH 
 is a much more challenging problem than in the Schwarzschild case.

\section{The regular Kerr  black hole with cosmological constant}

We will not follow here the same approach (the Newman-Janis algorithm \cite{Newman-Janis}) used by Bambi and Modesto in \cite{Bambi}, but rather we will employ
the so-called Synge $g$-method: assume $g_{\mu\nu}$, calculate and interpret $T_{\mu\nu}$. For an early application of
Synge $g$-method to the problem
of the interior of the Kerr black hole, see \cite{Synge}. Our main goal is to extend Bambi and Modesto solutions for the
case where a cosmological constant $\Lambda$ is present, and for other mass functions as well. We envisage two possible coordinate systems to explore here. The first 
one is related the so-called Kerr-Schild ansatz 
with cosmological constant
(see, for instance, \cite{Gibbons})
\begin{equation}
ds^{2}=ds_{\Lambda}^{2}+H\left(l_{\mu}dy^{\mu}\right)^{2},\label{Kerr-Schild}
\end{equation}
where 
$ds_{\Lambda}^{2}$ is a pure anti-de Sitter (AdS, $\Lambda <0$) or de Sitter (dS, $\Lambda>0$) metric, $H$ is a 
smooth function, and $l_{\mu}$ stands for a null vector. By introducing the 
 $(\tau,r,x=\cos \theta,\varphi)$ spheroidal coordinates, Eq. (\ref{Kerr-Schild})
can be decomposed as
\begin{equation}
ds_{\Lambda}^{2}=-\frac{(1-\frac{\Lambda}{3}r^{2})\Delta_{x}}{\Xi}d\tau^{2}+\frac{\Sigma}{(1-\frac{\Lambda}{3}r^{2})(r^{2}+a^{2})}dr^{2}+\frac{\Sigma}{(1-x^2)\Delta_{x}}dx^{2}+\frac{(r^{2}+a^{2})(1-x^2)}{\Xi}d\varphi^{2}\label{ds_linha}
\end{equation}
 and
\begin{equation}
l_{\mu}dy^{\mu}=\frac{\Delta_{x}}{\Xi}d\tau+\frac{\Sigma}{(1-\frac{\Lambda}{3}r^{2})(r^{2}+a^{2})}dr-\frac{a(1-x^2)}{\Xi}d\phi,\label{H-1}
\end{equation}
 where
\begin{equation}
\Delta_{x}=1+\frac{\Lambda}{3}a^{2}x^{2},\ \ \ \ \Sigma=r^{2}+a^{2}x^{2},\ \ \ \ \Xi=1+\frac{\Lambda}{3}a^{2}.\label{definitions}
\end{equation}
The constant $a$ will be later interpreted as the rotation parameter, but notice that it is present already in the pure
AdS or dS metric (\ref{ds_linha}) due to the use of  spheroidal coordinates. Notice that our construction requires $\Xi>0$, leading to
the restriction 
\begin{equation}
\label{cond}
\Lambda > -\frac{3}{a^2},
\end{equation}
and we will adopt this hypothesis hereafter. We will return to this point in the causal structure analysis in the next
section. 
We assume also rotational symmetry, and hence $H=H(r,x)$. A particularly convenient choice is
\begin{equation}
\label{H}
H(r,x)=\frac{2m(r)\ r}{\Sigma}. 
\end{equation}
For this case, with a $r$-dependent mass functions $m(r)$, one can introduce the usual  Boyer-Lindquist
coordinates $(t,r,x,\phi)$ by means of the following  
coordinates transformation
\begin{eqnarray}
d\tau&=&dt+\frac{\Sigma H}{(1-\frac{\Lambda}{3}r^{2})\Delta_{r}}dr,\label{transf_1} \\
d\varphi&=&d\phi-\frac{\Lambda}{3}adt+\frac{a\Sigma H}{(r^{2}+a^{2})\Delta_{r}}dr,\label{transf_2}
\end{eqnarray}
 where  
\begin{equation}
\Delta_{r}=(r^{2}+a^{2})\left(1-\frac{\Lambda}{3}r^{2}\right)-2rm(r).\label{Delta_r}
\end{equation}
The metric (\ref{Kerr-Schild})
will then take the Boyer-Lindquist form
\begin{eqnarray}
ds^{2} & = & -\frac{1}{\Sigma}\left(\Delta_{r}-\Delta_{x}a^{2}(1-x^2)\right)dt^{2}-\frac{2a}{\Xi\Sigma}\left[(r^{2}+a^{2})\Delta_{x}-\Delta_{r}\right](1-x^2) dtd\phi \nonumber \\
 & + & \frac{\Sigma}{\Delta_{r}}dr^{2}+\frac{\Sigma}{(1-x^2)\Delta_{x}}dx^{2}+\frac{1}{\Xi^{2}\Sigma}\left[(r^{2}+a^{2})^{2}\Delta_{x}-\Delta_{r}a^{2}(1-x^2)\right](1-x^2) d\phi^{2},\label{Metrica_Boyer-Lindquist}
\end{eqnarray}
which is much more convenient for the analysis of the causal structure of the
spacetime, and we will adopt it hereafter. We will come back in the last section to the case of possible mass functions
of the type $m=m(r,x)$, for which the 
above coordinate transformations are not defined. 

For $r$-dependent mass functions $m(r)$, 
the curvature scalar for the metric (\ref{Kerr-Schild}) or (\ref{Metrica_Boyer-Lindquist}) reads  simply
  \begin{equation}
 \label{scalar}
 R = 2 \frac{rm''(r) + 2m'(r)}{r^2+a^2x^2} + 4\Lambda,
 \end{equation}
 and, in spite of its simplicity, this expression retracts rather well the behavior of the singularities of the spacetime. For generic $m(r)$, we have a singularity at $x=0$ (equatorial plane) for $r\to 0$. The situation is equivalent to the divergence of the  Kretschmann scalar on a ``ring'' for the usual Kerr spacetime (constant $m$, $\Lambda=0$). In order to avoid such singularities for non constant $m$, the numerator of (\ref{scalar}) must vanish as $r^\alpha$ for small $r$, with $\alpha\ge 2$, which is satisfied for any mass function behaving as $m(r)\approx M_0 \left(\frac{r}{r_0}\right)^3$ for $r\to 0$. Moreover, for mass functions of this type, we have an ``effective'' de Sitter core (\ref{desitter}) on the equatorial plane ($x=0$) and
 the corresponding  Kretschmann scalar will read
\begin{equation}
\label{ket}
K = 96\left(\frac{M_0}{r_0^3}\right)^2\frac{r^4(r^8 + 4a^2x^2r^6 + 11a^4x^4 r^4 - 2a^6x^6 r^2 + 6a^8x^8)}{(r^2+a^2x^2)^6}
+32 \frac{M_0}{r_0^3} \frac{\Lambda r^2}{r^2+a^2x^2}
+\frac{8}{3}\Lambda^2,
\end{equation}
from where one can deduce straightly all the cases considered by Bambi and Modesto in \cite{Bambi} for $\Lambda=0$. The inclusion of the cosmological constant $\Lambda$ does not alter their conclusions about the avoidance of the central singularity.

 In order to analyse the matter content associated with (\ref{Metrica_Boyer-Lindquist}), we introduce the 
usual orthonormal tetrads \cite{Bardeen}
\begin{equation}
e^{(a)}_{\mu}=\left(\begin{array}{cccc}
\sqrt{\mp(g_{tt}-\Omega g_{t\phi})}& 0 & 0 & 0\\
0 & \sqrt{\pm g_{rr}} & 0 & 0\\
0 & 0 & \sqrt{g_{xx}} & 0\\
{g_{t\phi}}/{\sqrt{g_{\phi\phi}}} & 0 & 0 & \sqrt{g_{\phi\phi}}
\end{array}\right),\label{Matrix}
\end{equation}
which corresponds to the standard locally non-rotating frame, with 
 $\Omega = \frac{g_{t\phi}}{g_{\phi\phi}}$ being interpreted as the angular velocity of the BH. 
Notice that (\ref{cond}) suffices to assure  $g_{xx}>0$ and, since
\begin{equation}
g_{\phi\phi} = \frac{(r^2+a^2)(r^2+a^2x^2)\left(1+\frac{\Lambda}{3}a^2 \right) + 2a^2rm(r)(1-x^2)}{\left(1+\frac{\Lambda}{3}a^2 \right)^2(r^2+a^2x^2)},
\end{equation}
it also assures  a regular and positive $g_{\phi\phi}$. Moreover, from 
\begin{equation}
g_{tt}-\Omega g_{t\phi} =
-\frac{1}{3}\frac{(r^2+a^2x^2)\left(1+\frac{\Lambda}{3}a^2\right)\left[(r^2+a^2)\left(1-\frac{\Lambda}{3}r^2\right) -2rm(r)\right]}
{(r^2+a^2)(r^2+a^2x^2)\left(1+\frac{\Lambda}{3}a^2\right) + 2a^2rm(r)(1-x^2)}
\end{equation}
we see that the condition (\ref{cond}) also assures that $g_{tt}-\Omega g_{t\phi}$ does not diverge and has a opposite  sign of
\begin{equation}
g_{rr} = \frac{r^2 +a^2x^2}{(r^2+a^2)\left(1-\frac{\Lambda}{3}r^2\right) -2rm(r)}.
\end{equation}
Finally, 
 the signs in (\ref{Matrix}) must be
 selected in accordance  with  the considered region. The spacetime (\ref{Metrica_Boyer-Lindquist}) has generically two or three horizons (see next section) located at the zeros of $g^{rr}=g_{rr}^{-1}$. The innermost corresponds to a Cauchy horizon, and since we are considering BH solutions, a event horizon will be necessarily present. The region outside the event horizon and the region inside the Cauchy horizon correspond to the choice $(-,+)$, respectively, in the components
 $e^{(0)}_{0}$ and $e^{(1)}_{1}$. In this case, $e^{(0)}_{\mu}$ is timelike. On the other hand, the choice $(+,-)$ corresponds to the regions contained between the Cauchy and the event horizons, where  $e^{(1)}_{\mu}$ will be timelike. Our main argument here is based on the behavior of the energy momentum tensor near the origin and, hence, inside the Cauchy horizon. The expressions for the components of the energy-momentum tensor in the
 orthonormal tetrads frame $T^{(a)(b)} = \frac{1}{8\pi}e^{(a)}_{\mu}e^{(b)}_{\nu} G^{\mu\nu}$ are rather cumbersome, but they simplify
 considerably for $x=\pm 1$ (the ``poles'' along the rotation axis). In particular, for $x=\pm 1$, $T^{(a)(b)}$ is diagonal. One can check   
 that $T^{(0)(3)}$ does vanish for $x=\pm 1$ by a direct calculation, but this result could also be advanced from the fact that smooth ``tangential'' flows as the ones corresponding to  $T^{(0)(3)}$ must vanish along the symmetry axis. The non vanishing  $T^{(a)(b)}$ components for $x=\pm 1$ in the region outside the event horizon or inside the Cauchy horizon read simply 
\begin{eqnarray}
T^{(0)(0)}  &=& \frac{2r^2m'(r)}{8\pi (r^2+a^2)^2} + \frac{\Lambda}{8\pi}  = - T^{(1)(1)}, \\
T^{(2)(2)}  &=& -\frac{2a^2m'(r) + r(r^2+a^2)m''(r)}{8\pi (r^2+a^2)^2} - \frac{\Lambda}{8\pi} =  T^{(3)(3)} .
\end{eqnarray}
For the region inside the Cauchy horizon, $e^{(0)}_{\mu}$ is timelike and the WEC on the ``poles'' reads $T^{(0)(0)}\ge 0$ and
$T^{(0)(0)}+T^{(i)(i)} \ge 0$, $i=1\dots 3$. In the present case,
the condition for $i=2$ or $3$  
 requires $m'(r)\le 0$ near the origin for $a\ne 0$, 
a  rather unnatural requirement for any mass function in this context. In fact, for our paradigmatic case $m(r)\propto r^3$ one
has
\begin{equation}
T^{(0)(0)} + T^{(2)(2)} = T^{(0)(0)} + T^{(3)(3)}\propto -\frac{12r^2a^2}{(r^2+a^2)^2}
\end{equation}
for $r\approx 0$ and $x=\pm 1$, from where one can see that the the violation of WEC cannot be prevented if $a\ne 0$, irrespective
of the value of $\Lambda$ and the details of $m(r)$ far from the origin.

Notice that for $r \gg r_0$, one has
\begin{equation}
\label{R}
R \approx 
 4\Lambda + \frac{6M_0}{r_0^3} \frac{1-q}{1+ \frac{a^2x^2}{r^2}}  \left(\frac{r_0}{r}\right)^{q+3}
\end{equation}
for mass functions of the type (\ref{massfunc}) with $q>0$. We see from Eq. (\ref{R}) that the larger the value of $q$, the faster 
the solution converges  to the vacuum solution for large $r$, as one would indeed expect.
  The $q=1$ case corresponds to a Reissner-Nordstrom-like solution for which the energy-momentum tensor is traceless and, hence, only the cosmological constant counts for the Ricci scalar $R$. From (\ref{R}), one can estimate, for instance, the deviations from the usual Kerr solution. For a BH with $r_0 = \varepsilon M_0$, $\varepsilon<1$, the curvature deviations in the external region ($r>M_0$) have an  upper bound
  given by   $\varepsilon^{q+1}/M_0^2$. The exterior regions of a regular rotating and a Kerr BH can be almost indistinguishable for high values of $q$.

\section{spacetime causal structure of the regular solutions}

From the metric in the Boyer-Lindquist coordinates (\ref{Metrica_Boyer-Lindquist}),
we can find out the roots of $g^{rr}$, which will provide the radii of the horizons. The pertinent equation is $\Delta_{r}=0$,
which leads to
\begin{equation}
F(r)\equiv\frac{r^{2}+a^{2}}{2r}\left(1-\frac{\Lambda}{3}r^{2}\right) = m(r). 
\label{Delta_r=0}
\end{equation}
For $a\ne 0$, the left-handed side of (\ref{Delta_r=0}) behaves as $a^2/r$ for $r\approx 0$. On the other hand, for $\Lambda\ne 0$, it goes
as $-\Lambda r^4/3$
for large $r$.  Since the metric in the Boyer-Lindquist coordinates (\ref{Metrica_Boyer-Lindquist}) is independent of $t$
and $\phi$, we have two explicit Killing vectors $\xi_{t}=\frac{\partial}{\partial t}$ and  $\xi_{\phi}=\frac{\partial}{\partial\phi}$.
Since $g_{\phi\phi}>0$ by hypothesis due to the condition (\ref{cond}), $\xi_{\phi}$ is spacelike everywhere. For $\xi_{t}$, we
have 
\begin{equation}
|\xi_{t}|^2 =  g_{tt}=-
\frac{(r^2+ a^2x^2)\left(1-\frac{\Lambda}{3}(r^2+a^2(1-x^2)) \right) - 2rm(r)}{r^2+ a^2x^2}.
\label{Killing_norm1}
\end{equation}
A Killing horizon corresponds to a surface with a null type tangent Killing vector. 
Thus, one can locate Killing horizons by setting (\ref{Killing_norm1})  
to zero, and the pertinent equation is similar to (\ref{Delta_r=0}), namely
\begin{equation}
F_x(r)\equiv\frac{r^{2}+a^{2}x^2}{2r}\left(1-\frac{\Lambda}{3}(r^2+a^2(1-x^2))\right) = m(r). 
\label{Delta_rx=0}
\end{equation}
Notice that
\begin{equation}
F(r)-F_x(r) =\frac{ a^2(1-x^2)}{2r}\left(1+\frac{\Lambda}{3}a^2x^2 \right),
\end{equation}
and from (\ref{cond}), we have $F(r)\ge F_x(r) $. 
 The cases of AdS and dS are qualitatively different and we will treat them separately. Notice, however, that for $x=\pm 1$,
the Killing and event horizons coincide ($F(r)=F_x(r)$). The situation is identical to the Kerr solution. On the other hand, outside the symmetry
axis, the Killing and ordinary horizons do not coincide, giving origin to the ergoregions as in the usual Kerr solution.
 
\subsection{Asymptotically AdS case}

For the AdS case ($\Lambda <0$),  $F(r)$ has a global minimum located at
\begin{equation}
\label{rmin}
r_{\rm min}^2 = \frac{1}{6\Lambda}
\left(3-a^2\Lambda  - \sqrt{
\left(3- a^2\Lambda \right)^2 - 36 a^2 \Lambda}
\right).
\end{equation}
The existence of a black hole here requires two zeros for (\ref{Delta_r=0}), the inner (Cauchy, $r=r_{-}$) and the outer (event, $r=r_{+}$) 
horizons. A sufficient condition for
this is $F(r_{\rm min}) < m(r_{\rm min})$. It is not a surprise that certain combinations of parameters do not correspond effectively 
to black holes, a similar behavior is observed already for the simplest cases with $a=\Lambda=0$ \cite{Hayward}. 
The algebraic expressions for the roots are quite involved and we will  omitted them. Fig. 2 depicts some typical cases for the Cauchy and
event horizons. 
\begin{figure}
\begin{centering}
\includegraphics[scale=0.41]{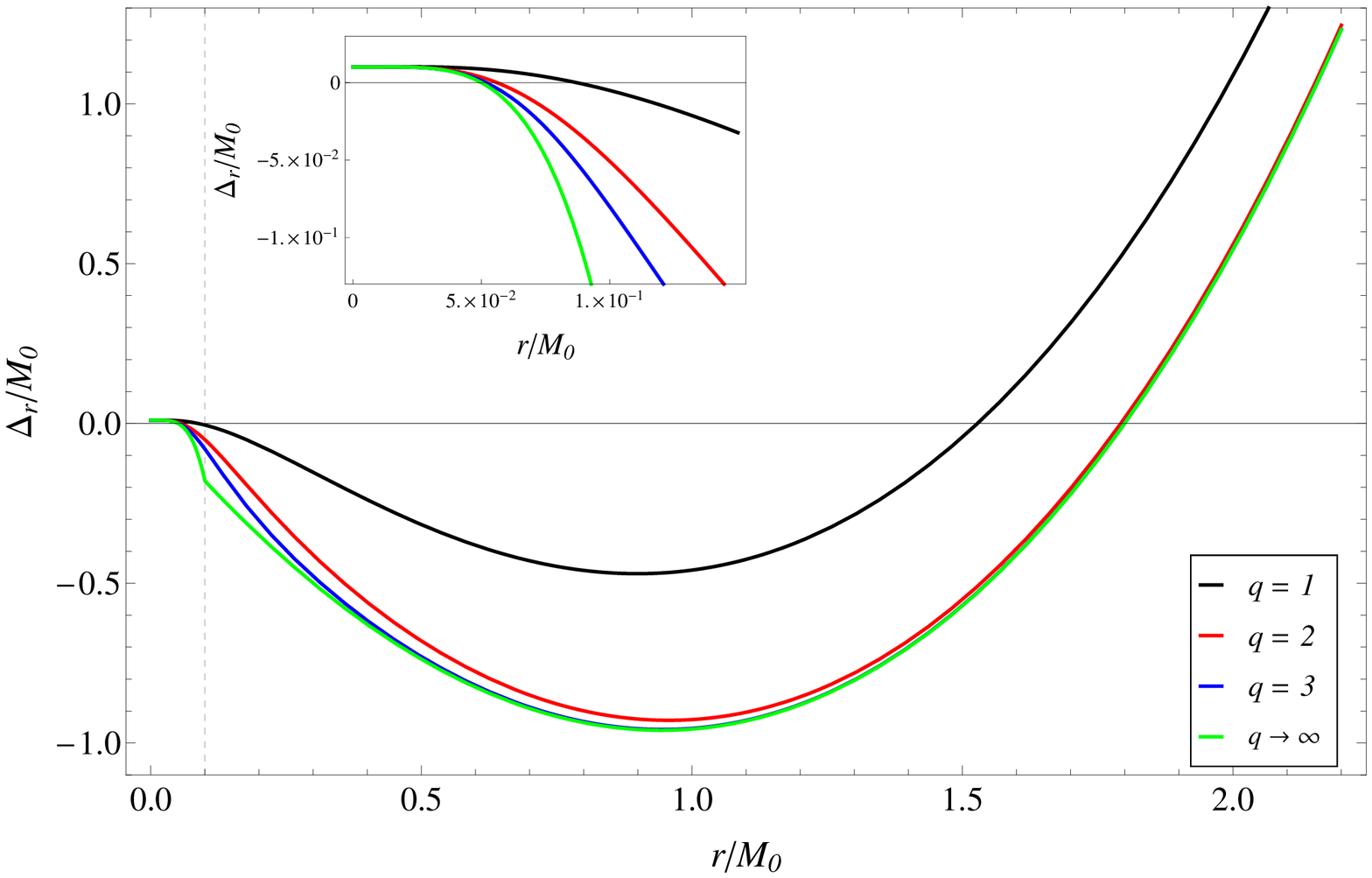}\includegraphics[scale=0.398]{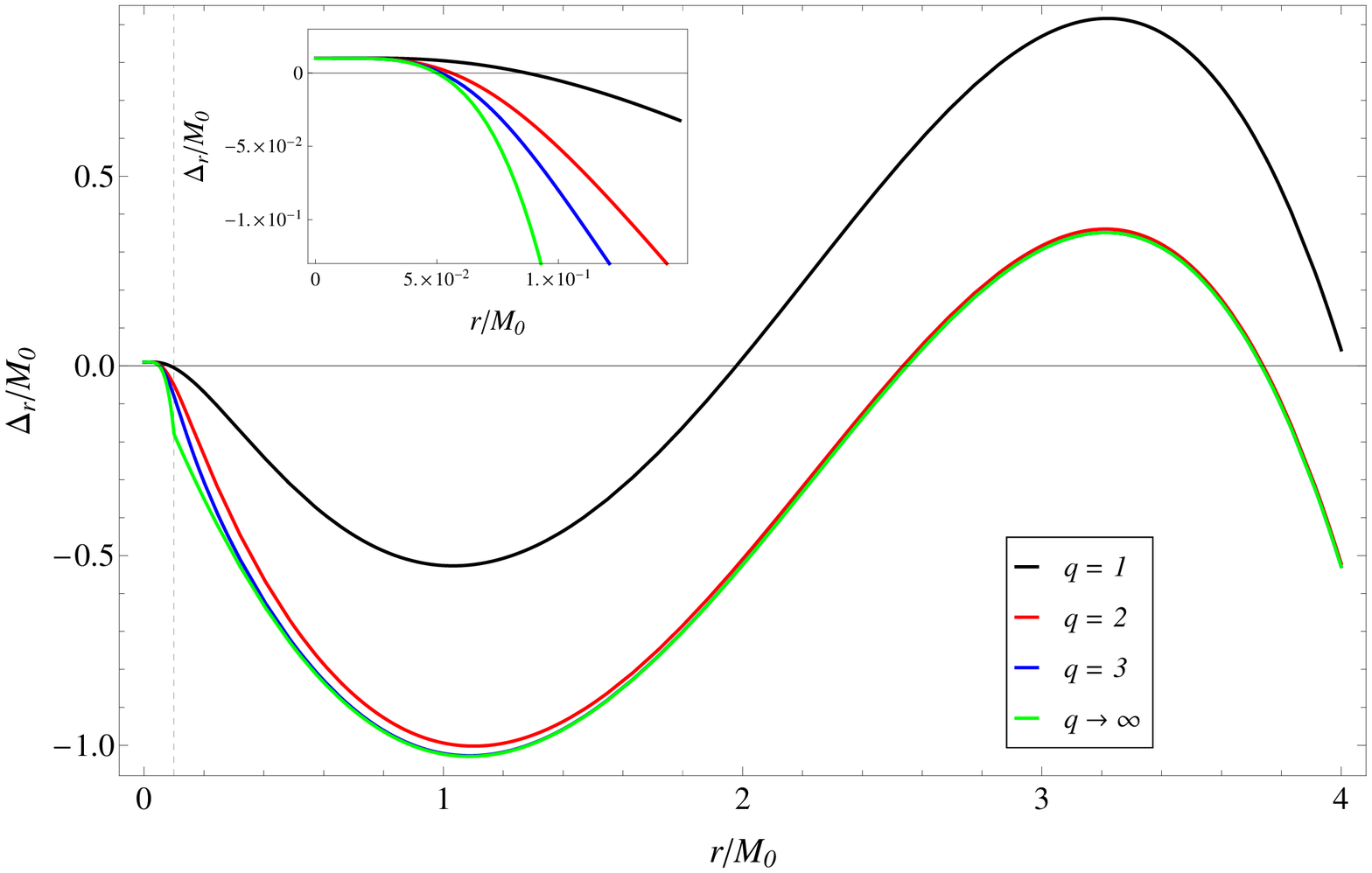}
\par\end{centering}
\caption{The horizons corresponding to the zeros of $\Delta_r$. Left: AdS case ($\Lambda < 0$), right: dS case ($\Lambda>0$). The dashed line indicates $r=r_0$. In these graphics we have used $p=3$, $ r_0/M_0 = a/M_0 = \pm\Lambda M_0^2 = 10^{-1}$.}
\end{figure}
Since $F(r)\ge F_x(r)$, the condition $F(r_{\rm min}) < m(r_{\rm min})$ also assures two roots for 
 (\ref{Killing_norm1}), which will correspond to the 
the Killing horizons $r=S_{-}$ and $r=S_{+}.$ For $x^2 \ne 1$,
we can devide the space-time
structure in five regions 
\begin{equation}
0<S_{-}<r_{-}<r_{+}<S_{+}<\infty.\label{AdS_region}
\end{equation}
This situation is depicted in Fig. 3. 
The region between $r_{+}$ and $S_{+}$ is the ergosphere, with the same properties of the usual esgosphere in the Kerr solution. In the
present case, however, we also have an interior ergosphere, which corresponds to the region limited by $S_-$ and $r_-$.

\begin{figure}
\begin{centering}
\includegraphics[scale=0.52]{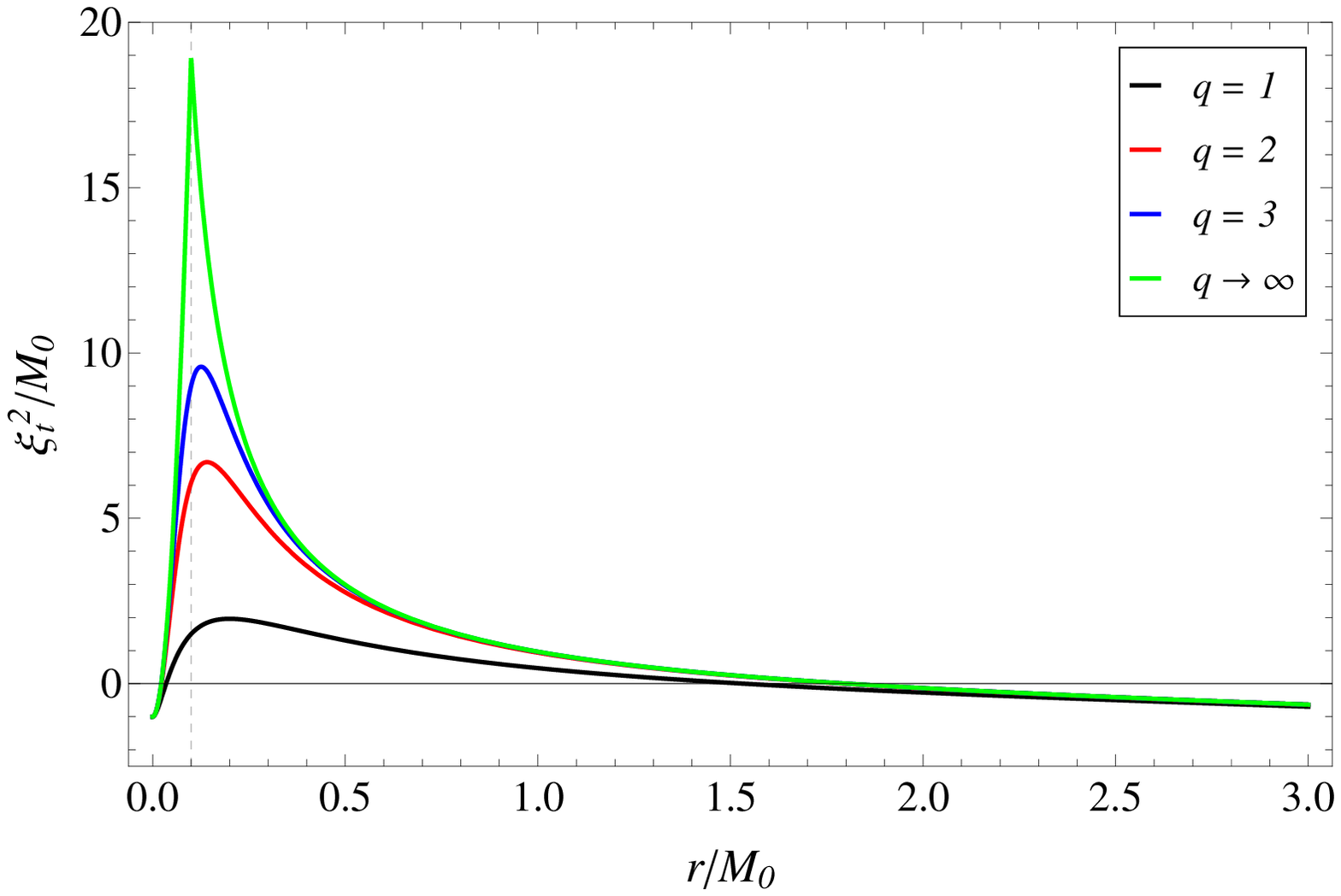}\includegraphics[scale=0.522]{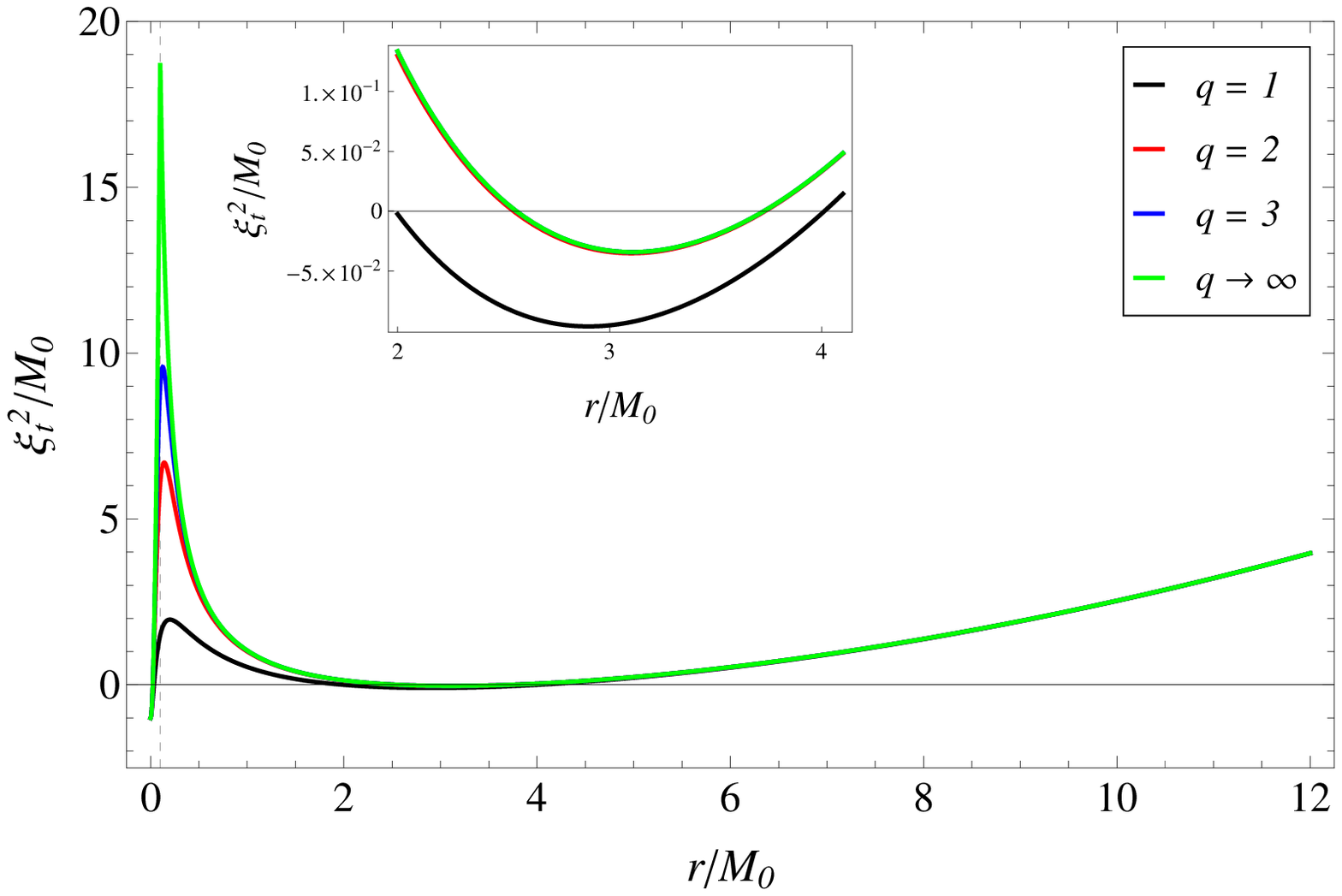}
\par\end{centering}

\caption{The Killing horizons correspondingo to the zeros of (\ref{Killing_norm1}), on the equatorial plane ($x=0$). Left: AdS case ($\Lambda < 0$), right: dS case ($\Lambda>0$). The dashed line indicates $r=r_0$. In these graphics we have used $p=3$, $ r_0/M_0 = a/M_0 = \pm\Lambda M_0^2 = 10^{-1}$.}

\end{figure}

\subsection{Asymptotically dS case}

The dS case ($\Lambda >0$) is more involved. The function $F(r)$ can have up to two critical points. Besides $r_{\rm min}$ given
by (\ref{rmin}), we have
also 
\begin{equation}
r_{\rm max}^2 = \frac{1}{6\Lambda}
\left(3-a^2\Lambda  + \sqrt{
\left(3- a^2\Lambda \right)^2 - 36 a^2 \Lambda}
\right).
\end{equation}
In order to have $r_{\rm min}$ and $r_{\rm max}$ properly defined, one needs $\Lambda < 3(7-4\sqrt{3})/a^2$. This condition, in addition
to $F(r_{\rm min}) < m(r_{\rm min})$ and $F(r_{\rm max}) > m(r_{\rm max})$, is sufficient to guarantee three zeros for (\ref{Delta_r=0}),
which corresponds to the inner $(r=r_{-})$, event $(r=r_{+})$, and cosmological $(r=r_{c})$ horizon. The situation where the cosmological and event horizon coincide corresponds to a rotating Nariai solution, see \cite{Nariai}. We will focus here only the situations containing black holes and, hence, we assume that all the
necessary conditions are met. 

For $\Lambda >0$ we can have also up to three Killing horizons, namely $S_{-},S_{i}$, and $S_{+}$. The spacetime can be divided 
into six regions for $x^2\ne 1$
\begin{equation}
0<S_{-}<r_{-}<r_{+}<S_{i}<S_{+}<r_{c}.\label{dS_region}
\end{equation}
The internal ergosphere is limited by $S_-$ and $r_-$, as in the AdS case. The external ergosphere consist effectively in two regions,
namely those ones limited by $r_+$ and $S_i$, and by $S_+$ and $r_c$. These regions are disjointed for $x^2\ne 1$.

\section{Final remarks}

We have here extended the recent work of Bambi and Modesto \cite{Bambi} where 
regular rotating BHs were introduced. Our solutions accommodate a cosmological constant $\Lambda$
and we have also introduced a more general mass function. We have shown that  
the black hole rotation will unavoidably lead to the violation of the weak energy condition (WEC) for any physically 
reasonable mass function. Despite of the violations of WEC, solutions as those ones introduced by
Bambi and Modesto and extended here are important not only from a phenomenological point of view, but it could also
contribute to the study of possible violations of the cosmic censorship conjecture  in quasi-extremal
black holes \cite{ccc}. These points are now under investigation. 

We finish by noticing that the case of 
 a $r$-dependent rotation parameter $a$ in the Boyer Lindquist
metric
(\ref{Metrica_Boyer-Lindquist}),
as discussed in  \cite{Bambi}, will give
origin to nonvanishing shear components (namely $T^{rx}$) in the energy momentum tensor, challenging the physical interpretation of the
matter content of such solutions,  as advanced early in the work \cite{av}. 
 The same occurs if one allows 
a mass function of the type $m(r,x)$. In this case, moreover, one cannot
obtain a Boyer Lindquist
metric from the Kerr-Schild ansatz since the coordinate transformations (\ref{transf_1})-(\ref{transf_2}) are not properly defined, leading
to an extra ambiguity: the solutions with $m(r,x)$ of the form (\ref{Kerr-Schild}) and (\ref{Metrica_Boyer-Lindquist})
are inequivalent, and both correspond to energy momentum tensors with nonvanishing shear components. It is not clear how to
interpret physically BH solutions with $r$-dependent rotation parameter $a$ and/or with mass functions of the type $m(r,x)$.

\begin{acknowledgments}
This work was supported by Funda\c c\~ao de Amparo \`a Pesquisa do Estado
de S\~ao Paulo (FAPESP), Brazil (Grant 2013/03798-3). AS is partially supported by CNPq (Brazil).
\end{acknowledgments}


\begin{thebibliography}{References}

\bibitem{sing}P.S. Joshi, {\em Spacetime singularities}, 
to appear in the Springer Handbook of Spacetime (Eds A. Ashtekar and V. Petkov),
[arXiv:1311.0449].

\bibitem{Novello}M. Novello, S. E. Perez Bergliaffa. Phys. Rept. \textbf{463}, 127 (2008) [arXiv:0802.1634].

\bibitem{Ansoldi} S. Ansoldi, {\em Spherical black holes with regular center: a review of existing models including a recent realization with Gaussian sources}, [arXiv:0802.0330].

\bibitem{Heusler} M. Heusler, {\em Black holes uniqueness theorems}, Cambridge University Press (1996).

\bibitem{Hayward}S. A. Hayward, Phys. Rev. Lett. \textbf{96}, 031103 (2006) {[}arXiv:gr-qc/0506126{]}.


\bibitem{Sakharov} A.D. Sakharov, JETP {\bf 22}, 241 (1966); E.B. Gliner, JETP
{\bf 22}, 378 (1966).

\bibitem{Markov} M.A. Markov, JETP Lett. {\bf 36}, 265 (1982); V.P. Frolov,
M.A. Markov, and V.F. Mukhanov, Phys. Rev. D {\bf 41}, 383
(1990); V.F. Mukhanov and R. Brandenberger, Phys. Rev.
Lett. {\bf 68}, 1969 (1992); R. Brandenberger, V.F. Mukhanov,
and A. Sornborger, Phys. Rev. D {\bf 48}, 1629 (1993).




\bibitem{Bambi}C. Bambi, L. Modesto, Phys. Lett. B \textbf{721},
329-334 (2013) {[}arXiv:gr-qc/1302.6075{]}.


\bibitem{Newman-Janis}E. T. Newman and A. I. Janis, J. Math. Phys.
\textbf{6}, 915 (1965).


\bibitem{Hawking}S. W. Hawking, C. J. Hunter, and M. M. Taylor-Robinson,
Phys. Rev. D \textbf{59}, 064005 (1999) {[}arXiv:hep-th/9811056{]}.


\bibitem{Synge} W.C. Hernandez, Jr., Phys. Rev. {\bf 159}, 1070 (1967).

\bibitem{Gibbons}G. W. Gibbons, H. Lu, D. N. Page, and C. N. Pope,
J Geom. and Phys. \textbf{53}, 49 (2005) {[}arXiv:hep-th/0404008{]}.

\bibitem{Bardeen}J. M. Bardeen, W. H. Press and S. A. Teukolsky.
The Astrophy. J \textbf{178}, 347 (1972).

\bibitem{Nariai} D. Anninos and  T. Anous, JHEP {\bf 1008}, 131 (2010) [arXiv:1002.1717].



\bibitem{ccc} M. Richartz and A. Saa, Phys. Rev. D {\bf 78}, 081503 (2008) [arXiv:0804.3921]; 
G.E.A. Matsas, M. Richartz, A. Saa, A.R.R. da Silva, and D.A.T. Vanzella, Phys. Rev. D {\bf 79}, 101502 (2009) [arXiv:0905.1077];
A. Saa and R. Santarelli, Phys. Rev. D {\bf 84}, 027501 (2011) [arXiv:1105.3950]; M. Richartz and A. Saa, 
Phys. Rev. D {\bf 84}, 104021 (2011) [arXiv:1109.3364].


\bibitem{av}M. Murenbeeld and J. R. Trollope, Phys. Rev. D{\bf 1}, 3220 (1970). 



\end{thebibliography}
\end{document}